\documentclass[a4paper]{jpconf}
\usepackage{graphicx}
\usepackage{amssymb,amsmath,array}
\usepackage{subfigure}
\usepackage[vcentering,dvips]{geometry}

\begin{document}

\title{System size and beam energy dependence of azimuthal anisotropy from PHENIX}

\author{Michael Issah (for the PHENIX Collaboration)}
\address{Department of Physics and Astronomy,
Vanderbilt University,
1807 Station B,
Nashville TN 37235, USA}
\ead{michael.issah@vanderbilt.edu}

\begin{abstract}

We present azimuthal anisotropy measurements in Au+Au and Cu+Cu collisions at $\sqrt{s_{NN}}$ 
= 62.4 and 200 GeV. Comparison between reaction plane and cumulant $v_2$ measurements in Au+Au 
collisions at $\sqrt{s_{NN}}$ = 200 GeV show that non-flow contributions, originating mainly from 
jets, influence the extracted $v_2$ for $p_T$ $\gtrsim$ 3.5 GeV/c. Number of constituent quark (NCQ) 
scaling of $v_2$, when studied as a function of transverse kinetic energy $KE_T$, is seen to hold for Au+Au 
collisions at $\sqrt{s_{NN}}$ = 62.4 and 200 GeV and for Cu+Cu collisions at $\sqrt{s_{NN}}$ 
= 200 GeV for $KE_{T}$ $\lesssim$ 1 GeV/c. Differential hexadecupole flow $v_4$ seems to exhibit 
scaling with integral $v_2$ for centrality $\le$ 40$\%$ 
as has been observed for differential $v_2$.

\end{abstract}

\section{Introduction}
\label{intro}

Experimental measurements at the Relativistic Heavy Ion Collider (RHIC) have indicated the 
creation of a deconfined phase of quarks and gluons~\cite{Adcox:2004mh}. The azimuthal 
anisotropy of particles emitted suggests that the high-energy density matter rapidly 
thermalizes and its evolution is driven by large pressure gradients~\cite{Gyulassy:2004zy,Heinz:2005ja}. Azimuthal 
correlation measurements in Au+Au collisions at RHIC have been shown to consist 
of a mixture of jet and harmonic contributions~\cite{Adler:2002tq}. The azimuthal anisotropy 
extracted from these correlations are thus the sum of contributions from collective flow and 
jet-like correlations. The latter constitute one source of non-flow correlations, especially at high
 $p_T$, and one would like to study how they vary with $p_T$ and centrality. It has been reported
that constituent quark scaling of $v_2$ as a function of $KE_T$ or $m_T - m$ is observed in 
Au+Au collisions at $\sqrt{s_{NN}}$= 200 GeV and that differential $v_2$ scales with integral 
$v_2$ for both Au+Au and Cu+Cu collisions at $\sqrt{s_{NN}}$= 200 GeV~\cite{Adare:2006ti} .
These observations have implications on the hydrodynamic description of the medium as well as 
the pertinent degrees of freedom in the flowing matter~\cite{Adare:2006ti}.
PHENIX has measured the $v_2$ of charged pions, kaons and protons in Au+Au collisions at $\sqrt{s_{NN}}$ 
= 62.4 GeV and  Cu+Cu collisions at $\sqrt{s_{NN}}$= 200 GeV. It is interesting to study 
whether NCQ scaling holds for these two colliding systems and if the scaling of 
differential $v_2$ with integral $v_2$ also works for differential $v_4$.

\section{Data analysis}
\label{analysis} 

The results presented here use data obtained by the PHENIX collaboration for Au+Au collisions 
at $\sqrt{s_{NN}}$= 62.4 and 200 GeV and Cu+Cu collisions at $\sqrt{s_{NN}}$= 200 GeV. The 
azimuthal anisotropy relevant to this work were measured using two different methods: 
the reaction plane~\cite{Ollitrault:1993ba,Poskanzer:1998yz} and the cumulant~\cite{Borghini:2001vi} 
method. The reaction plane method measures the correlations between the azimuthal 
angle of the particles detected in the central arms of the PHENIX detector and the azimuth of the reaction 
plane obtained with the two Beam-Beam Counters (BBC) located at $\mid{\eta}\mid \sim 3 - 3.9$~\cite{Adler:2003kt}. 
The cumulant method extracts $v_2$ from multiparticle correlations of particles detected in the 
PHENIX Central arms~\cite{Adler:2004cj}.
The large $\eta$ separation between the BBC's and the central arms of PHENIX means that the 
extracted anisotropy is much less affected by possible non-flow contributions, especially those 
due to jets~\cite{Adler:2003kt,:2008cq}. The differential second-order cumulant 
$v_2$ is determined with respect to the reconstructed $v_2$ integrated over $p_T$. Monte Carlo 
simulations have been carried out to ensure that the azimuthal anisotropy is reconstructed 
faithfully with the PHENIX detector acceptance~\cite{Borghini:2001vi,Adler:2004cj}. The 
particles used to reconstruct the differential flow are excluded from those used in the 
integral flow reconstruction to avoid autocorrelations. The second order cumulant $v_2\{2\}$ is extracted 
from two-particle correlations and thus includes correlations from particles originating from jet 
fragmentation. Hence, the second order cumulant $v_2$ is influenced by non-flow correlations 
from jets as discussed in~\cite{Borghini:2001vi}. The results presented in the next section show how 
these jet correlations affect the azimuthal anisotropy in Au+Au collisions at $\sqrt{s_{NN}}$= 200 GeV.

\section{Results}
\label{results}
\subsection{Comparison between  $v_2$ using different methods of measurement}
\label{comparison}

\begin{figure}[!htb]
\begin{center}
\includegraphics[width=0.7\linewidth]{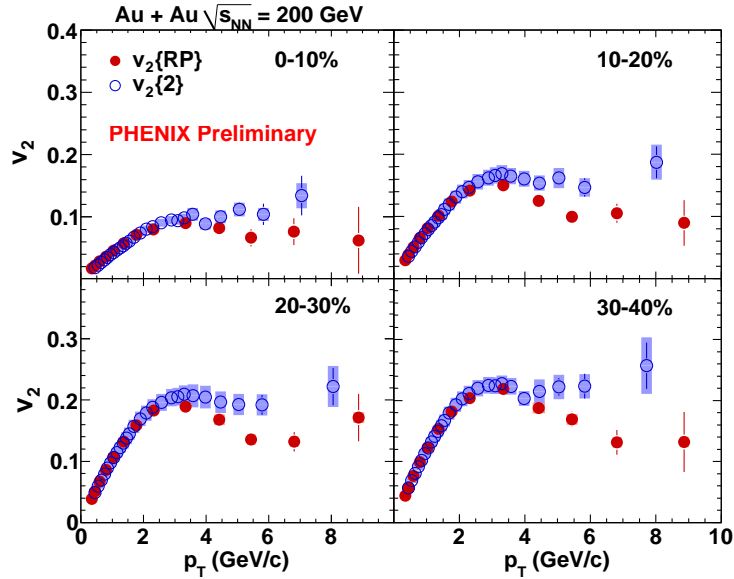}
\vspace*{-0.5cm}
\caption{Comparison between $v_2$ from cumulant and reaction plane (RP) methods for different 
centralities in Au+Au at $\sqrt{s_{NN}}$ = 200 GeV. The closed and open circles, labelled
$v_2\{RP\}$ and $v_2\{2\}$, are the cumulant and reaction plane results respectively.}
\label{v2cumrp}
\end{center}
\end{figure}
Fig.~\ref{v2cumrp} shows the differential $v_2$ in  Au+Au at $\sqrt{s_{NN}}$ = 200 GeV as a 
function of $p_T$ from the reaction plane and the cumulant methods for several centralities. 
The measured $v_2$ from the two methods agree well 
up to $p_T$ $\approx$ 3.5 GeV/c but differ for higher $p_T$, with the cumulant $v_2$ being higher 
than the reaction plane $v_2$. This could be attributed to the influence of jets on the 
extracted $v_2$ which is more prominent in the cumulant method. One also notes that the same 
trend is observed at all centralities, even though the agreement between
the two sets of results holds up to higher $p_T$ for the 0-10$\%$ centrality bin.

\subsection{Number of constituent quark scaling}
\label{scaling}
\begin{figure}[!htb]
\begin{center}
\includegraphics[width=0.85\linewidth]{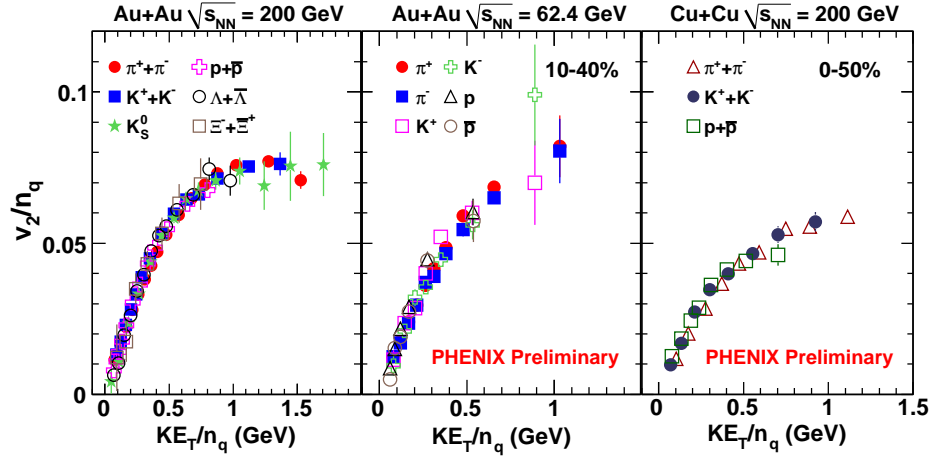}
\caption{$v_2$/$n_q$ as a function of $KE_T$/$n_q$ for charged pions, kaons and protons 
in minimum bias Au+Au collisions at $\sqrt{s_{NN}}$ = 200 GeV (left panel), in Au+Au 
collisions at $\sqrt{s_{NN}}$ = 62.4 GeV for centrality 10~-~40$\%$ (middle panel) and 
Cu+Cu at $\sqrt{s_{NN}}$ = 200 GeV for centrality 0~-~50$\%$ (right panel).}
\label{v2ket}
\end{center}
\end{figure}

Number of constituent quark scaling of $v_2$ in Au+Au at $\sqrt{s_{NN}}$ = 200 GeV works 
well when studied as a function of transverse kinetic energy, as shown in the left panel
of Fig.~\ref{v2ket}~\cite{Adare:2006ti}. This scaling scheme has been extended to the $v_2$ 
of positive and negative pions, kaons and protons measured in Au+Au collisions at $\sqrt{s_{NN}}$ 
= 62.4 GeV and Cu+Cu collisions at $\sqrt{s_{NN}}$ = 200 GeV. The middle panel of Fig.~\ref{v2ket} 
shows that the constituent quark scaling of $v_2$ works well for $\pi^{\pm}$, $K^{\pm}$ and 
($\bar{p}$)$p$ in Au+Au collisions at $\sqrt{s_{NN}}$ = 62.4 GeV. The same good scaling of 
$v_2$ is observed for the smaller Cu+Cu system at $\sqrt{s_{NN}}$ = 200 GeV, as shown in 
the right panel of Fig.~\ref{v2ket}. These observations suggest that partonic degrees of 
freedom exist in the matter formed at lower beam energy and in the smaller colliding system.

\subsection{$v_4$ of unidentified particles}
\label{v4} 
\begin{figure}[!htb]
\begin{center}
\includegraphics[width=0.65\linewidth]{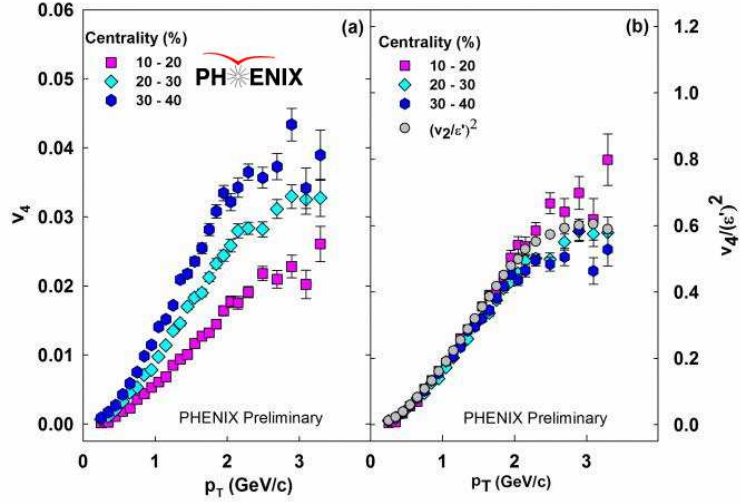}
\caption{Left panel: Differential $v_4$ of unidentified hadrons as a function of $p_T$ (left panel) for 
different centralities. Right panel: $v_4$/${\epsilon^{'}}^2$ as a function of $p_T$, 
where $\epsilon^{'}$ represents $v_2$ integrated over $p_T$. For the sake of comparison,
the open circles represent differential $v_2$ squared scaled by the square of the integral $v_2$.}
\label{v4pt}
\end{center}
\end{figure}

PHENIX has recorded high statistics Au+Au collisions during the 7th year of data-taking,
enabling the detailed study of higher harmonics of azimuthal anisotropy as a function
of $p_T$ and centrality. It has been reported that $v_2$ scales with the integral flow, which 
has been shown to be proportional to the eccentricity~\cite{Adare:2006ti}. The use
of the integral flow scaling of differential $v_2$ has been used to study the 
thermalization of the matter at RHIC~\cite{Adare:2006ti}. Since, in general, $v_n$ $\sim$ $v_2^{n/2}$, which 
yields $v_4$ $\sim$ $v_2^{2}$ for $n =$ 4, differential $v_4$ would be expected to scale with 
the square of the differential $v_2$ and the scaling of differential $v_4$ with integral $v_2$
could be investigated. The left panel of Fig.~\ref{v4pt} shows the $p_T$ dependence of $v_4$ 
of unidentified particles for three centrality bins in the range 10-40~$\%$ and the right panel shows 
$v_4$, scaled by the square of the $p_T$-integrated $v_2$ for the corresponding
centrality (labelled $\epsilon^{'}$ in the figure), as a function of $p_T$.
The latter plot shows that $v_4$ scales with the square of integral $v_2$ for 
centrality $\leq$ 40~$\%$ and $p_T$ $\lesssim$  2.0 GeV/c. By plotting  $({v_2/\epsilon^{'}})^2$ on 
the same plot (open circles) one observes a good agreement between differential $v_4$ and $v_2^2$, when 
they are both scaled by the square of integral $v_2$. 

\section{Summary}
\label{summary}
In summary, we have presented results of azimuthal anisotropy from different methods at RHIC. 
They show that non-flow correlations from jets may influence the measured 
anisotropy for $p_T$ $\gtrsim$ 3.5 GeV/c. Constituent quark scaling is observed to 
hold in Au+Au at $\sqrt{s_{NN}}$ = 62.4 and 200 GeV and Cu+Cu at $\sqrt{s_{NN}}$ = 200 GeV. 
This suggests that partonic degrees of freedom are manifest at the lower RHIC beam energy and in 
the smaller Cu+Cu colliding system. The differential hexadecupole flow $v_4$ in Au+Au at  
$\sqrt{s_{NN}}$ = 200 GeV is observed to scale with the integrated $v_2$ for centrality $<$ 
40$\%$ and $p_T$ $\lesssim$ 2.0 GeV/c.

\end{document}